# Observation of Rydberg exciton polaritons and their condensate in a perovskite cavity

Wei Bao[1†], Xiaoze Liu[1†], Fei Xue[2,3†], Fan Zheng[4], Renjie Tao[1], Siqi Wang[1], Yang Xia[1], Mervin Zhao[1], Jeongmin Kim[1], Sui Yang[1], Quanwei Li[1], Ying Wang[1], Yuan Wang[1], Lin-Wang Wang[4], Allan MacDonald[2*], Xiang Zhang[1, 5*]

[1]NSF Nanoscale Science and Engineering Center, University of California, Berkeley, California 94720, USA.

[2]Department of Physics, University of Texas at Austin, Austin, Texas 78712, USA

[3]Institute for Research in Electronics and Applied Physics & Maryland Nanocenter, University of Maryland, College Park, MD 20742, USA

[4]Materials Sciences Division, Lawrence Berkeley National Laboratory, Berkeley, California 94720, USA.

[5]Faculties of Sciences and Engineering, University of Hong Kong, Hong Kong, China

† These authors contributed equally to this work.
*Correspondence to: xiang@berkeley.edu or macd@physics.utexas.edu

**The condensation of half-light half-matter exciton polaritons in semiconductor optical cavities is a striking example of macroscopic quantum coherence in a solid state platform. Quantum coherence is possible only when there are strong interactions between the exciton polaritons provided by their excitonic constituents. Rydberg excitons with high principle value exhibit strong dipole-dipole interactions in cold atoms. However, polaritons with the excitonic constituent that is an excited state, namely Rydberg exciton polaritons (REPs), have not yet been experimentally observed. Here, for**

**the first time, we observe the formation of REPs in a single crystal CsPbBr$_3$ perovskite cavity without any external fields. These polaritons exhibit strong nonlinear behavior that leads to a coherent polariton condensate with a prominent blue shift. Furthermore, the REPs in CsPbBr$_3$ are highly anisotropic and have a large extinction ratio, arising from the perovskite's orthorhombic crystal structure. Our observation not only sheds light on the importance of many-body physics in coherent polariton systems involving higher-order excited states, but also paves the way for exploring these coherent interactions for solid state quantum optical information processing.**

Solid state cavity quantum electrodynamics (CQED) delivers extraordinary control of light matter interactions in various photonic structures(1). Beyond simply modifying the photonic density of states in the weak coupling regime, CQED also enables the formation of new hybrid light-matter quasiparticles called cavity polaritons(2,3). In semiconductor microcavities (MC), cavity polaritons are created by strong coupling between excitons and photons when the coupling rate is faster than the dissipation rates of both constituents. These bosonic quasiparticles possess a small effective mass (~$10^{-4}$ electron mass)$^2$ from their photonic component and inherit strong interactions from their excitonic component. The combination leads to rich quantum optical phenomena, such as polariton condensation, superfluidity, and quantum vortices, that are similar to those seen in cold atom Bose-Einstein condensation (BEC), but at much elevated temperatures(4–8).

Polariton condensation relies on strong nonlinear polariton interactions via their matter constituent, and is characterized by a macroscopic coherent condensate of strongly interacting bosonic particles in a non-equilibrium state(4,6,9). The exciton-exciton interactions emerge from underlying Coulomb interactions, and strongly depend on the dielectric environment and the exciton radius(4). Currently, the relatively delocalized Wannier-Mott excitons in inorganic semiconductors have formed polariton condensates at densities that are much lower than the exciton's Mott density(10–12), while the tightly-bound Frenkel excitons in organic semiconductors have reached polariton condensation at high exciton densities around the exciton saturation density(13,14). This difference is due to the larger exciton-exciton interaction strength for Wannier-Mott excitons (typical size 3-10 nm) than for Frenkel excitons (typical size <1 nm)(15), leading to orders of magnitude stronger interactions in Wannier-Mott excitons. Importantly, there are also excited states of Wannier-Mott excitons that are predicted to provide stronger dipole-dipole interaction strength in the framework of hydrogen-like Rydberg states(16). Polaritons with excitonic constituent of excited states, namely Rydberg exciton polaritons (REPs), have not been naturally observed, due to weak oscillator strength of the excitonic excited states in most optically active semiconductors.

Recently, the emerging lead halide perovskites with Rydberg exciton series(17,18) have provided a high quality optoelectronic platform that does not require sophisticated lattice-matched growth(19,20). The group of lead halide perovskite semiconductors have bulk excitons with exceptional optical properties(21–25), such as a sizeable exciton binding energy, tunable band gaps, high quantum yield, and Rydberg exciton series of strong oscillator strength without applying external magnetic fields(17,18,26). They are therefore excellent candidates for investigating exciton-polariton states and polariton condensation, and even for future quantum photonic circuits(7). Encouragingly, polariton lasing based on the ground exciton state has recently been demonstrated in a $CsPbCl_3$ microcavity(27). In this work, we demonstrate the formation of hybrid exciton polaritons in single crystal perovskite $CsPbBr_3$, including emerging REPs without external fields. More importantly, we show that Bose-Einstein condensation of polaritons is reached with a prominent blue shift and interesting mode competition that can be explained by our quasiequilibrium mean-field theory. In addition, these polaritons are anisotropic with a large extinction ratio driven by the anisotropy of the potential landscape in the perovskite's orthorhombic phase(28). This precise polarization control is a necessary prerequisite in quantum optical information processing. This work represents a major step in solid state quantum photonics systems, not only offering a unique platform for new quantum coherent many-body physics(29), but also opening a new door for solid state quantum photonic applications in communication and computing(30).

The metal halide perovskite $CsPbBr_3$ is selected as the exciton host. Compared to hybrid organic-inorganic halide perovskites, all-inorganic $CsPbBr_3$ exhibits superior chemical stability and emission efficiency(31). Fig. 1d shows typical absorption spectra of a $CsPbBr_3$ crystal on mica at 100K in its thermodynamically stable orthorhombic phase (see additional characterization in Materials and Methods; Fig. S1, S2 and S3)(28) with clear absorption peaks ($E_1$ and $E_2$), consistent with the recent studies(18,20). According to DFT in the orthorhombic phase (Fig. 1e), there is only one exciton formed in the band edges of $CsPbBr_3$, where all other electronic are separated by more than 1eV. The $E_1$ and $E_2$ peaks as the ground and first excited states (Rydberg series $n = 1, 2$) of the only exciton series are identified by the observed energy separation between the two absorption peaks, their temperature dependences (Fig. S2) and the reported exciton binding energy(26). The orthorhombic phase also results in the mutually orthogonal birefringence along the *a* and *b* crystalline axes (Fig. 1c, 1e).

To investigate the strong light-matter interactions in these excitonic states, we embedded the $CsPbBr_3$ microplate in a Fabry-Perot planar cavity (Fig. 1a), where the *a* and *b* axes are orthogonal along the in-plane surface. With the full cavity structure, the in-plane wavevector *k* of the cavity mode is also along this surface (Fig. 1a inset). As shown in Fig. 1b, the microplate transferred onto a DBR substrate has a uniform thickness of 416 nm (Fig. S1) and a typical square shape. Our cavity provides a quality factor in excess of a thousand, derived from the off-resonance cavity linewidth (Fig. S6) as well as polariton linewidth from Fig. 2. This high cavity quality assists the formation of a REP, owing to the sharp interface between the perovskite and the metal mirror (Fig. S1) as well as the reduced metal absorption losses at cryogenic temperature(32).

The coherent coupling of these states and cavity photons is revealed by *k*-space spectroscopy (Fig. S12 for more details), when the sample is first cooled down to 90K. The *k*-space characterization is carried out with selective linear polarization for both photoluminescence (PL) and reflectivity measurements (see Fig. 2). The PL measurements are taken using a non-resonant pump laser of 460 nm diagonally polarized between *a* and *b* axes (at a 45°, as shown in Fig. 1b), while the reflectivity measurements are carried out using a nonpolarized tungsten halogen white light source. When the detection polarization is set along the *a* axis, two dispersive modes are observed from both PL (Fig. 2a) and reflectivity (Fig. 2d), identified as newly formed polariton states. Although *k*-space reflectivity has lower color contrast (low reflectivity dips of less than 10%) than PL, they have clear one-to-one correspondence for these polariton states. Both of these polariton modes are distinctively flattened at larger emission angles, indicating that the two excitonic states are strongly coupled with the cavity modes. By applying a coupled oscillator model with the two excitonic states and a cavity mode at slightly positive detuning ($\Delta_a = E_{cav} - E_{a1} = 5$ meV), the lower and middle branches (labeled as LP$_a$ and MP$_a$, and overlaid with PL for better visibility) are consistent with the PL and reflectivity dispersion (Fig. 2a and 2d, see the model fitting analysis in Supplementary Note 1, additional supporting data in Fig S4 and Fig. S6, and the Hopfield coefficients in Fig. S10). The upper branch (UP$_a$ and UP$_b$) dominated by the photon mode component is not visible due to its weak oscillator strength and widely detuned cavity resonance, further aggravated in many cases by fast polariton relaxation(27). These two excitonic states corresponding to the $n = 1, 2$ states are confirmed to be in the strong coupling regime based on the temperature-dependent measurements (Fig. S5). The three-branch polariton dispersion holds up to 150K, below which the $n = 1, 2$ states can be resolved with large binding energy and oscillator strengths. The polariton dispersion becomes two-branch above 150K, beyond which only the ground $n = 1$ state can be resolved. Based on the measurements of a variety of samples, the red shift of the exciton energy levels relative to Fig. 1 is due to sample-to-sample variations, PL Stokes shifts, and temperature-dependent band edge shift (see more details in Fig. S2, S3; Materials and Methods), consistent with previous reports(13). Thus this observation implies coherent strong coupling between light and an exciton excited state without an external field for the formation of REP. The extracted energy splittings (37.4 meV between LP$_a$ and MP$_a$, and 29.6 meV between MP$_a$ and UP$_a$) are much larger than the polariton linewidth (~ 3 meV), demonstrating robust coherent coupling.

A similar PL dispersion has also been observed when setting the detection polarization along the *b* axis. By applying the same coupled oscillator model, the modes along the *b* axis are confirmed as strongly coupled

REPs (labeled as LP$_b$ and MP$_b$). Note that the polariton dispersion here is distinct from the *a* axis because of anisotropic refractive indices for the cavity modes, and different oscillator strengths along the two axes. When the detection polarization is along the *a-b* diagonal, the k-space PL map in Fig. 2c shows both sets of polarion dispersions linearly superimposed, suggesting that they are mutually orthogonal to each other. The independence of these polariton states is also seen in the PL intensities of each polariton branch (LP$_a$, MP$_a$, LP$_b$ and MP$_b$) in Fig. 2f with strong angular dependence (extinction ratio >50) along the *a* and *b* axes, respectively. Thus, these polaritons show extremely strong polarization anisotropy originating from the perovskite refractive indices.

The polaritons can reach the nonlinear polariton condensation regime at large population densities of these polaritons. A polariton condensate is a coherent ensemble of a finite density of particles in the lowest available polariton state, and can be described by a dissipative Bose-Einstein condensation model(4,6,7,11,33)(Supplementary Note 3). A condensate is possible at elevated temperatures (cryogenic temperature and above), due to the small effective mass (~$10^{-4}$ electron mass) of the hybrid light-matter particles and the strong interactions among them(4,6,7). The strong coupling regime of the REPs in Fig. 2 is observed in the linear regime with negligible interactions among them. At high carrier densities the interactions become significant, generating a stimulated nonlinear regime, and eventually a macroscopically coherent quantum condensation state.

Polariton condensation can also be observed by performing k-space PL measurements at various pump powers. Here, we select a different sample with a more positive cavity detuning ($\Delta_a = E_{cav} - E_{a1} = 25.5$ meV), where the excitonic fraction is significantly larger to generate strong exciton interactions(4,34)(See Fig. S10 for the Hopfield coefficients). The four k-space PL map panels in Fig.3a, labeled from left to right as 0.05P$_{th}$, 0.4P$_{th}$, P$_{th}$, and 1.4P$_{th}$, clearly show nonlinear threshold behavior with increasing pump power. In Fig.3 the threshold pump power P$_{th}$ ~ 6.8µJ/cm$^2$ pump pulse energy; the average P$_{th}$ in various samples is around 2.4 µJ/cm$^2$. See Fig. S11 for another similar example and a statistical summary in supplementary note 2). At each pump power, the detection polarization is set to be unpolarized so that both *a* and *b* polariton states can be observed. At pump power $0.05P_{th}$, the $LP_a$, $MP_a$, and $LP_b$ modes are clearly detected and in excellent agreement with the coupled oscillator model (solid curves). As the pump power reaches $0.4P_{th}$, the lower $LP_a$ and $LP_b$ branches dominate the PL spectrum, with the PL intensity of $LP_b$ increasing faster due to its higher excitonic fraction (see Supplementary Note 3 and Fig. S10 for further discussion of exciton fractions). It is worth noting that the original parameters for the Rydberg levels in the coupled oscillator model have to be blue shifted for these REPs, indicating an energy modification caused by the excitonic interactions(35,36). The effect is prominent until $P_{th}$. Importantly, the PL intensity of $LP_a$ increases dramatically, compared with $LP_b$. When the pump power is increased slightly over $P_{th}$ to 1.4 $P_{th}$, a full collapse of k-space PL dispersion into the bottom of $LP_a$, accompanied with a super-linear intensity increase and a large blue shift, suggesting that a condensate of these polaritons may have been reached. Moreover, the condensation process is also revealed by the lifetime data and statistics analysis (Fig. S7 and S8).

We perform further analysis to confirm polariton condensation. In Fig. 3b the PL intensity of $LP_a$ along the normal emission angle and the corresponding linewidth versus pump power are plotted on log-log and linear-log scales. The "S" shaped L-L curve of the PL intensity is divided into three regimes: the linear regime where polariton interactions are insignificant, a super-linear regime with a dramatic narrowing of linewidth due to the stimulated interactions between REPs, and the condensate regime in which most of REPs share the same lowest $LP_a$ state(4). Note here that the average exciton density of perovskite devices at $P_{th}$ is ~ 25 times lower than the exciton Mott density of the devices. This further justifies our interpretation of the non-equilibrium steady state as an exciton-polariton condensate rather than a normal photonic laser (see Supplementary Note 2 for more details).

In contrast to conventional polariton condensation, where only one polariton mode responds to the nonlinear interaction process(4,6,7), here strong exciton-exction interactions also lead to a unique condensation dynamics with multiple polariton modes involved. In Fig. 3c, the PL intensities of both $LP_a$ and $LP_b$ are plotted versus pump power. Below the threshold, the intensities of $LP_a$ and $LP_b$ increase with pump power independently. Due to the mutually orthogonal polarizations, these two modes barely crosstalk in the linear regime. This can be confirmed by the intensity ratio between $LP_a$ and $LP_b$ below threshold shown in the solid black spheres of Fig. 3c. When the pump power exceeds threshold, the increasing density of $LP_b$ does not contribute to the PL of $LP_b$, but directly interacts with other polaritons to establish the $LP_a$ condensate. This dynamic process is elaborated in our theory analysis based on dissipative Gross-Pitaevskii equations (GPEs) in which $LP_a$ is thermodynamically favored over $LP_b$ when the stimulated scattering reaches a more stable final-state (Supplementary Note 3). The final condensation on the $LP_a$ mode also suggests an efficient way to pin the polarization of the condensation, distinct from the stochastic polarization(37).

The strong interaction of the REPs in the nonlinear regime is also evident in the blue shifts of these polariton energies. This blue shift does not come from a heating effect at higher pump powers, as indicated by the similar shift under different pump conditions with significantly reduced heat (see control experimental data in Fig. S9). The PL peak positions are extracted as a function of pump power for both $LP_a$ and $LP_b$ (Fig. 3d). Below the threshold, both polariton modes show noticeable blue shifts due to small disorder effect and the reservoir repulsive exciton-polariton interactions even when the densities are not large(35,36,38). Above the threshold, repulsive polariton-polariton interactions become more prominent causing the REP modes to show strong blue shifts. Though the potential renormalization will blue shift the condensation of $LP_a$ as expected, the blue shift is quite significant while the $LP_b$ intensity is saturated here. This suggests that polaritons of the $LP_b$ modes actively contribute to the stimulated scattering process once the power is above the threshold (Fig. 3c, and Supplementary Note 3). We emphasize that the blue shift in both modes is almost 1/3 of the Rabi splitting, indicating very strong polariton interactions with the participation of Rydberg excitonic states. We have developed a theory model based on mean-field theory considering short-range exchange contributions and the dissipative GPEs to describe this large blue shift with ground state excitonic interactions (Supplementary Note 3 and 4, dot-dot-dash lines in Fig. 3d). Though exchange interaction between 2s excitons is surprisingly attractive rather than repulsive (details in the Supplementary Note 3), the 2s exciton fraction in the LP is significant smaller than 1s exciton which makes this negative contribution negligible. It is worth noting that a detailed theory with quantitative analysis including dipole-dipole interactions(16) and inter-exciton interactions (1s-2s interactions) and accurate experimental calibration of polariton density to get the better estimate of interaction strength are beyond the scope of the current work. This experimental observation of REP with enhanced interactions promises future explorations of Rydberg interactions in solid state systems.

In summary, we have surprisingly discovered REPs in a single crystal perovskite cavity, which enables coherent control of these fine quantum states. The intrinsic strong exciton interaction and optical birefringence in perovskite leads to the observation of polariton-condensation dynamics, which promises a robust macroscopically coherent state for quantum applications. This discovery presents a unique platform to study quantum coherent many-body physics, and potentially enables unprecedented manipulation of these Rydberg states by new means such as chemical composition engineering, structural phase control, and external gauge fields. Controlling the REP and its condensates not only adds new flavors on studying polariton lasing, superfluidity and vortices; but also holds great potential for important applications, such as communication, and quantum simulation.

**Acknowledgements** W.B. thanks Prof. E. Yablonovitch for valuable advice and discussion of the manuscript. W.B., X.L., R.T., S.W., Y.X., M.Z., J.K., S.Y., Q.L., Y.W., Y.W., and X.Z. acknowledge the support from the U.S. Office of Naval Research (ONR) MURI program (grant N00014-17-1-2588) and the National Science Foundation (NSF) under MRI Grant 1725335. F.X. acknowledges support under the Cooperative Research Agreement between the University of Maryland and the National Institute of Standards and Technology Physical Measurement Laboratory, Award 70NANB14H209, through the University of Maryland. F.Z., and L.-W.W. thank the Joint Center for Artificial Photosynthesis, a DOE Energy Innovation Hub, supported through the Office of Science of the U.S. Department of Energy under Award number DE-SC0004993. F.Z., and L.-W.W. use the resource of National Energy Research Scientific Computing center (NERSC) located in Lawrence Berkeley National Laboratory and the computational resource of the Oak Ridge Leadership Computing Facility at the Oak Ridge National Laboratory under the Innovative and Novel Computational Impact on Theory and Experiment project. The authors also acknowledge the facility support at the Molecular Foundry by the U.S. Department of Energy, Office of Science, Office of Basic Energy Sciences under Contract No. DE-AC02-05CH11231.

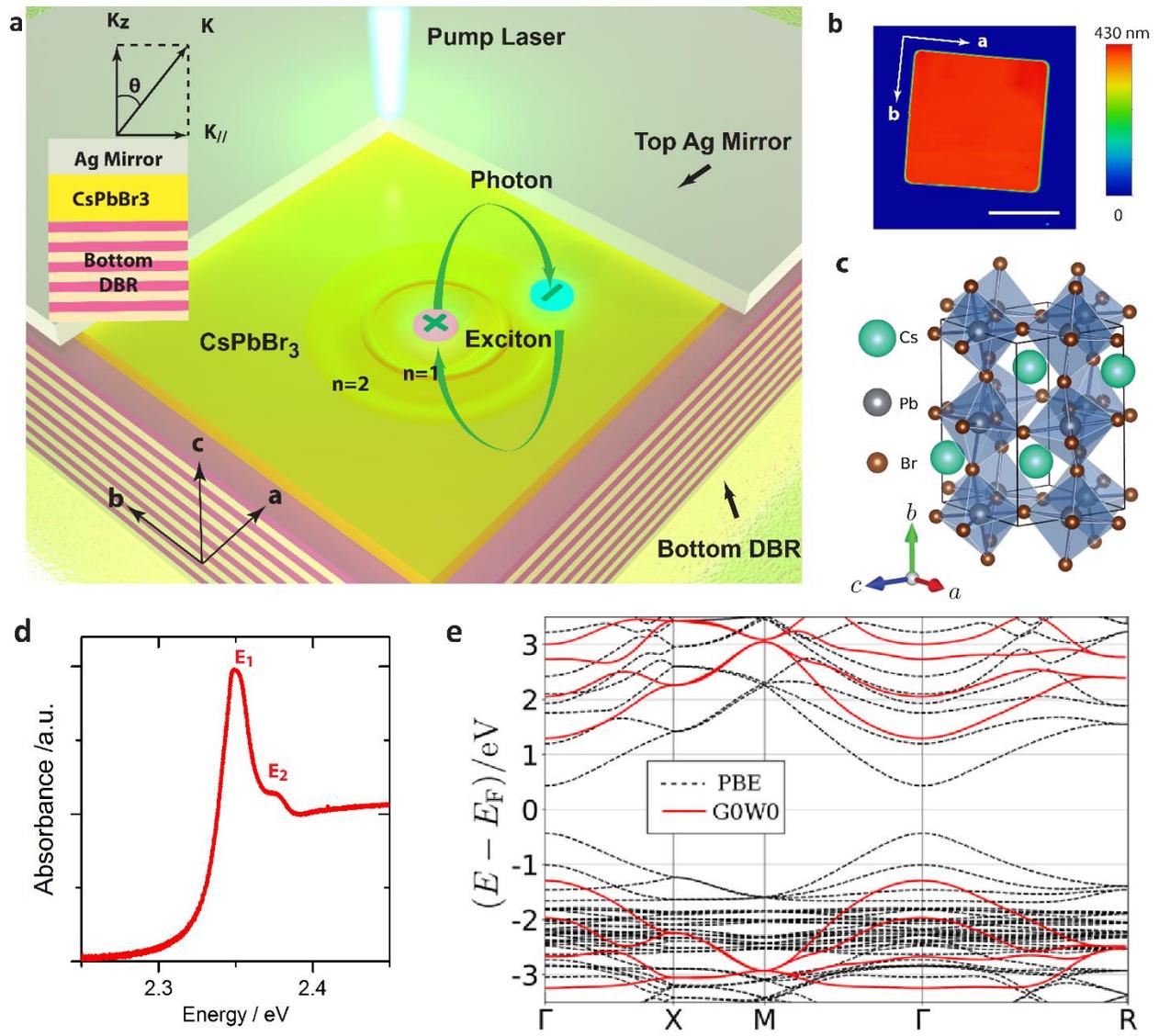

**Figure 1. Schematics of CsPbBr3 microcavity devices and materials characterization.** (a) The CsPbBr$_3$ microcavity is composed of a 16 pair SiO$_2$/Ta$_2$O$_5$ bottom distributed Bragg reflector (DBR), CVD grown CsPbBr$_3$ microplates with a thickness of 416 nm, and a 55 nm thick Ag top mirror. The crystal axes are also indicated. (b) Atomic force microscopy image of the uniform CsPbBr$_3$ square-shaped single crystal perovskite used in combination with the bottom DBR mirror in the experiments summarized in Fig.2. The crystal axes are also labeled. The scale bar corresponds to 10 µm. (c) The DFT calculated stable crystal structure of orthorhombic CsPbBr$_3$, with labeled *a*, *b*, and *c* crystalline axes. This structure results in almost identical refractive indices along the *a* and *c* axes, and a distinctly different refractive index along the *b* axis. (d) The polarization nonselective absorption spectrum of single crystal CsPbBr$_3$ film on mica at 100K. A prominent ground state E$_1$ exciton absorption peak is clearly shown along with the excited n = 2 Rydberg exciton $E_2$ state. **(e)** Calculated PBE and G0W0 band structures for orthorhombic $CsPbBr_3$. With the inclusion of spin-orbit coupling, the PBE calculated band gap is corrected to 2.5 eV by G0W0, agreeing well with the experiments. Importantly, unlike GaAs, $CsPbBr_3$ has no degenerate or nearby band states at conduction or valance band edges ($\Gamma$ point).

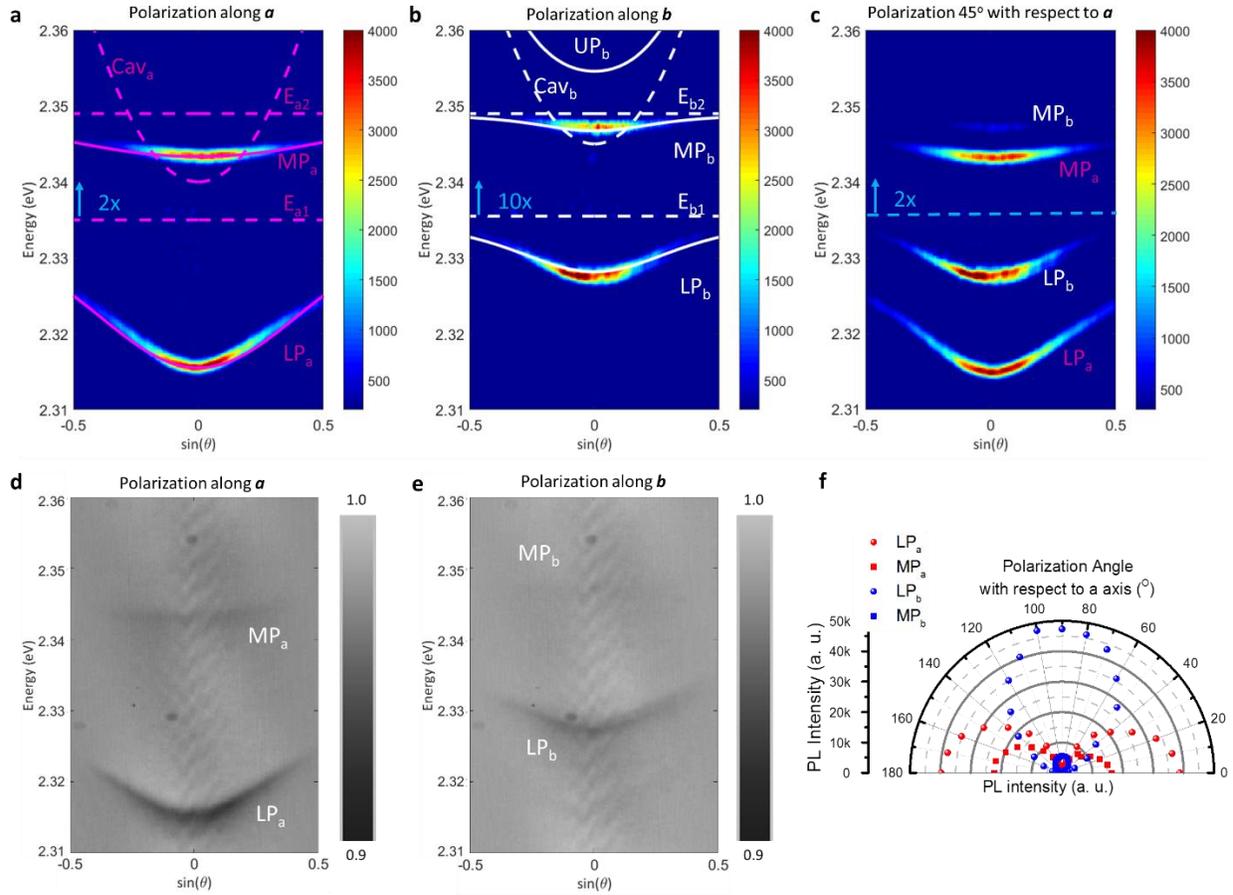

**Figure 2. K-space angle-resolved photoluminescence and white light reflectivity at 90 K.** The non-resonantly pumped (460 nm laser) PL map obtained by k-space spectroscopy with detection photon polarization **(a)** along crystal axis ***a***, **(b)** along crystal axis ***b***, and **(c)** 45° in between ***a*** and ***b*** axes (Fig. 1a. and 1b.). The intensity of the middle branch polariton PL is magnified by 2×, 10× and 2× in panel **a**, **b** and **c** respectively, due to its weak emission. The horizontal axis represents the sine function of the emission light slant angles θ relative to the z axis (Fig. 1a inset) and the vertical axis is the photon energy. Middle branch polariton $MP_a$ and $MP_b$ (better seen in **(c)**) are unambiguously formed due to the n = 2 exciton state. The polariton dispersion is fit using a coupled oscillator model. The exciton energy and photonic cavity mode ($Cav_a$ and $Cav_b$) before strong coupling (dashed line) and the fitted polariton dispersion (solid line) are overlaid with the PL map. These fine excitonic states and their polariton structures can only be observed at low temperatures (<150 K). At higher temperatures, the cavity samples transit from only one lower polariton branch to a broad PL peak (similar to bare exciton emission). The corresponding polarization selective white light reflectivity maps of the same sample **(d)** along crystal axis ***a***, **(e)** along crystal axis ***b***. The dispersion of k-space

reflectivity maps matches the PL dispersion fit very well. **(f)** The polarization dependence of polariton emission at normal angle ($\theta = 0°$). The extinction ratio of these two orthogonal lower branch emission modes is more than 50.

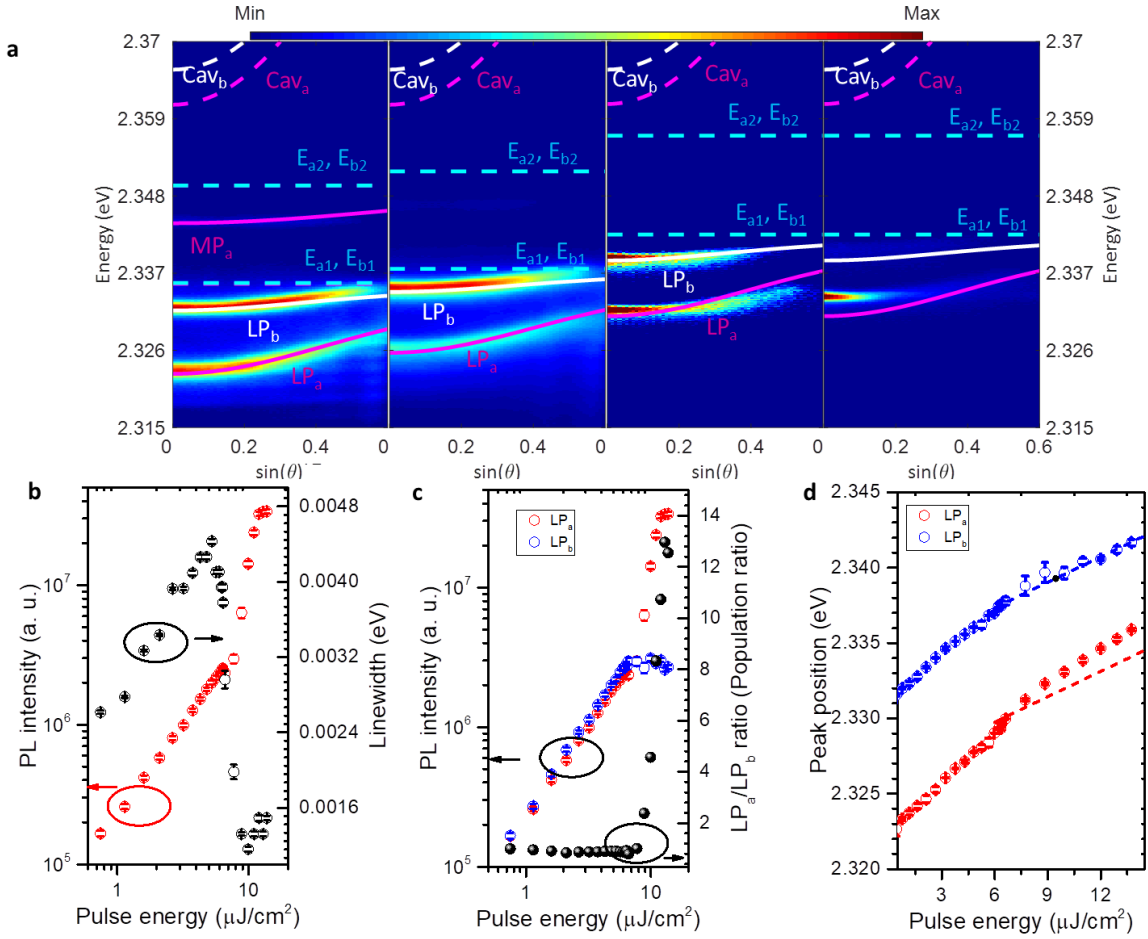

**Figure 3. Anomalous exciton-polariton condensate behavior at 55 K. (a)** k-space power dependent angle resolved PL map taken at 0.05 $P_{th}$, 0.4$P_{th}$, $P_{th}$ and 1.4$P_{th}$ (from left to right). The excitation is 460 nm light polarized along the ***a-b*** diagonal. The sample is slightly thinner and more positively detuned than in Fig. 2. The two sets of orthogonal Rydberg exciton-polariton modes are unambiguously identified and the polariton dispersions are fit using the same coupled oscillator model as in Fig. 2. The uncoupled exciton energy and photonic cavity mode dispersion (dashed line) and polariton dispersion fit (solid line) are overlaid with the PL map. The magenta color represents polarization mode along ***a*** axis while the white color represents the orthogonal polarization mode along ***b*** axis (Fig 1a. and 1c.). The 1.4$P_{th}$ panel shows the same fitt as at $P_{th}$ to emphasize the blue shift above the threshold. The small deviation in the high angle (sinθ) fitting of the polariton branch $LP_a$ and $LP_b$ at $P_{th}$ and 1.4$P_{th}$ is due to renormalization of the cavity mode at threshold  The polariton condensate experiences an anomalous condensation process in which the $LP_b$ shows a faster increase than the lower energy $LP_a$ state between the second and third panels. This is due to a stronger exciton interaction along the ***b*** axis. As the pump density gets close to the condensation density, the $LP_a$ finally experiences a superlinear increase with stimulated scattering to the lowest $LP_a$ state while $LP_b$ shows no further increase. **(b)** Log-log plot of integrated PL intensity of $LP_a$ mode at $\theta = 0°$ and full-width at half-maximum (FWHM) of $LP_a$ mode at $\theta = 0°$ versus pump power. Nonlinearity and linewidth narrowing of the polariton mode is observed as the excitation intensity exceeds the condensation threshold. Fitting error bars from the data processing are shown in **(b)**, **(c)** and **(d)**. **(c)** Log-log plot of both $LP_a$ mode (red dot) and $LP_b$ mode (blue dot) at $\theta = 0°$. PL intensity and the ratio of the two modes versus pump power. **(d)** PL peak position of both $LP_a$ mode (red dot) and $LP_b$ mode (blue dot) at $\theta = 0°$ versus pump power. A strong blue shift of polariton modes below the threshold is observed due to the strong exciton interactions and potential system disorder(39). After the threshold of condensation, a prominent blue shift in both $LP_a$ and $LP_b$ mode results from the polariton-polariton interaction and the polariton-reservoir interaction. The theory predicted blue shift contributed from 1s-exciton-resulted polariton interaction is plot in red and blue dot-dot-dash line for guidance. The experimental observed value is larger than the estimation from pure 1s exciton interaction.

# Supplementary Information for

# Observation of Rydberg exciton polaritons and their condensate in a perovskite cavity

Wei Bao[1†], Xiaoze Liu[1†], Fei Xue[2,3†], Fan Zheng[4], Renjie Tao[1], Siqi Wang[1], Yang Xia[1], Mervin Zhao[1], Jeongmin Kim[1], Sui Yang[1], Quanwei Li[1], Ying Wang[1], Yuan Wang[1], Lin-Wang Wang[4], Allan MacDonald[2*], Xiang Zhang[1, 5*]

[1]NSF Nanoscale Science and Engineering Center, University of California, Berkeley, California 94720, USA.

[2]Department of Physics, University of Texas at Austin, Austin, Texas 78712, USA

[3]Institute for Research in Electronics and Applied Physics & Maryland Nanocenter, University of Maryland, College Park, MD 20742, USA

[4]Materials Sciences Division, Lawrence Berkeley National Laboratory, Berkeley, California 94720, USA.

[5]Faculties of Sciences and Engineering, University of Hong Kong, Hong Kong, China

† These authors contributed equally to this work.

* Correspondence to: xiang@berkeley.edu or macd@physics.utexas.edu

**This PDF file includes:**

    Materials and Methods
    Supplementary Note 1-4
    Supplementary Figures S1 to S13
    References

**Materials and Methods**

CVD All-inorganic CsPbBr$_3$ microplate growth

Single crystal CsPbBr$_3$ microplates are grown on 150 μm thick high quality muscovite mica substrates via CVD, modified from the references (1, 2). Prior to growth, the substrates are cleaned by scotch tape exfoliation. The growth surfaces are placed face-down on the top of a quartz crucible that contained a fine power mixture with 20 mg of CsBr (Sigma Aldrich 99.999% purity) and 30mg of PbBr$_2$ (Sigma Aldrich 99.999% purity). The system is pumped down to below 50 mTorr while purged with ultrahigh purity Ar gas at a flow rate of 500 sccm for 2 min. Then the pump is shut down and tube pressure is maintained at ambient pressure with an Ar flow rate of 30 sccm. The system is firstly heated to 400 °C within 24 min, from 400 °C to 500 °C within 4 min and 30 sccm Ar , held at 500 °C for 20 min, and stopped afterwards. When the temperature reaches ~200 °C, the furnace is completely opened to achieve rapid cooling. During the cooling stage, Ar is continually supplied to the furnace tube at a 30 sccm to prevent sample degradation. The samples are characterized with photoluminescence (PL) and energy dispersive X-ray spectroscopy (EDS), and show a main PL peak around 525 ± 5 nm at room temperature and a stoichiometry ratio of CsPbBr$_3$.

Sample preparation

The metal-DBR cavities adopted in the experiments combine ease of fabrication with adequate quality factors. A single-crystal silicon (100) wafer with deep etched markers (35 um deep) is first pre-cleaned by a Piranha solution. Then the bottom 16 pairs of SiO$_2$ and Ta$_2$O$_5$ distributed Bragg reflector (DBR) is deposited on top via ion beam sputtering (Veeco IBS) at a temperature of 120 ºC with a pressure of $< 5\times10^{-5}$ Torr. The sputtering rate is set to be ~0.1 nm/s to achieve ultrahigh flatness and > 99.95% reflectivity. The CsPbBr$_3$ can be directly transferred from the mica substrate to the bottom DBR using a PDMS stamp (Dow Corning Sylgard 184 Silicone Encapsulant kit), which is cured at room temperature and a mixture ratio of 7 to 1(3). Since the CsPbBr$_3$ microplates (typically ~400 nm and can form 3λ cavities) would have their quantum yield degraded by direct e-beam evaporation of a top Ag mirror, we thermally evaporated 2 nm Al (3 samples, one in Fig.

2 and one in Fig. 3) or 10 nm $Sb_2O_3$ (one in Fig. S6 and can preserve PL quantum better) as a seeding layer (at 0.01 nm/s) followed by 5 nm $Al_2O_3$ atomic layer deposition (ALD) (Oxford Instrument). Lastly, the sample is coated with 55 nm 99.99% purity silver using e-beam evaporation (CHA solution) with a pressure of $< 5 \times 10^{-7}$ Torr. The evaporation rate is set to be ~1 nm/s to avoid oxidation during evaporation.

Optical characterization

The k-space measurements are carried out in a home-built confocal setup attached with a spectrometer and an EMCCD camera (Andor spectrometer). More details of the k-space measurements are discussed in Fig. S12. This confocal setup is also integrated with a Janis open cycle cryostat which can be cooled down to liquid helium temperature. The long working distance objective lens (NIKON Plan Fluor ELWD 40x, 0.6 NA) was used with an objective correction collar to see through the cryostat window. The diameter of the pump beam focused by this objective is around 4 μm. The reflectivity is measured with a collimated white lamp source (Ocean optics HL-2000 Tungsten Halogen) while the PL is measured with a fs pulsed laser. The pulsed laser is extracted by an optical parametric oscillator (Inspire, Spectra Physics) pumped by a mode-locked Ti:Sapphire oscillator. The laser pulse width is around 200 fs and the repetition rate is 80 MHz. The absorption spectrum was measured using a home-built UV-VIS-IR transmission setup, which utilizes the same collimated Ocean optics white lamp source, Andor spectrometer and Andor EMCCD camera.

**Supplementary Note 1: Three branch polariton dispersion fits**

The polariton dispersion can be modelled using a simplified coupled oscillator model in which linewidths are neglected, since they are quite small compared to the coupling strengths (4, 5):

$$\begin{pmatrix} E_{cav}(\theta) & V_1 & V_2 \\ V_1 & E_1 & 0 \\ V_2 & 0 & E_2 \end{pmatrix} \begin{pmatrix} \alpha \\ \beta \\ \gamma \end{pmatrix} = E \begin{pmatrix} \alpha \\ \beta \\ \gamma \end{pmatrix}$$

Here the cavity mode $E_{cav}(\theta) = E_{ph} / \sqrt{1-(\sin\theta/n_{eff})^2}$ where $E_{ph}$ is the photon cut-off energy and and $n_{eff}$ is an effective refractive index $n_{eff}$ of the cavity layer. $E_1, E_2$ are the Rydberg exciton energies ($E_{a1}, E_{a2}$ are for the series along the *a* axis, and $E_{b1}, E_{b2}$ along the *b* axis). $E$ is an eigenvalue which corresponds to the energy of the polariton modes, and $\alpha, \beta,$ and $\gamma$ define the corresponding eigenvector. Note that $\alpha, \beta,$ and $\gamma$ satisfy $|\alpha|^2 + |\beta|^2 + |\gamma|^2 = 1$, and represent the Hopfield coefficients of the cavity photon and the Rydberg exciton states for each polariton state. $V_1, V_2$ are the coupling strengths between the photon and the two exciton states. The Rabi splitting at detuning $\Delta_1 = E_{cav} - E_1 = 0$ can be expressed as $\hbar\Omega_1 \approx 2V_1$, and that at detuning $\Delta_2 = E_{cav} - E_2 = 0$ as $\hbar\Omega_2 \approx 2V_2$. Note that this model focuses on plotting out the energy dispersion, but does not describe the linewidths, which are the broadening of the measured results. For the sample demonstrated in Fig. 2, the fitted dispersions are in excellent agreement with experimental data. The Rydberg exciton energies are degenerate along *a* and *b* axes, as they should be, and are taken as $E_{a1} = E_{b1} = 2.3355$ eV and $E_{a2} = E_{b2} = 2.3505$ eV. Along the *a* axis, the cavity modes are modelled by $E_{ph\_a} = 2.3330$ eV and $n_{eff\_a} = 3.80$. The Rabi splittings are then calculated as $\hbar\Omega_{a1} = 32.0$ meV and $\hbar\Omega_{a2} = 24.0$ meV. Along the *b* axis, the cavity modes are modelled by $E_{ph\_b} = 2.3457$ eV and $n_{eff\_b} = 3.75$. The Rabi splittings are then calculated as $\hbar\Omega_{b1} = 21.2$ meV and $\hbar\Omega_{b2} = 12.0$ meV. This anisotropy originates from the orthorhombic crystal structure. The dispersion fitting depends on the effective indices of the cavity layer. For simplicity, these values are picked from the median values of the indices in specific photon energy ranges for Fig. 2, where the indices have some variations as shown in Fig. S6. Because of the small variations of the real indices, the measured results may have some slight shift from the fitting model at some photon energy ranges. These shifts are less than 1 meV, which still show consistent fitting with measured results. The fitting of Fig. 3 as discussed later is also similar.

In the sample of Fig. 3, the detuning is much more positive so that the lower branches have more than 65% excitonic fraction, implying strong Rydberg interactions for polariton condensation. At the pump power of 0.05 $P_{th}$ ($P_{th}$ is the threshold pump power) in the linear regime, the Rydberg energies along *a* and *b* axes are very similar to those in Fig. 2. We find that $E_{a1} = E_{b1} = 2.3355$ eV and $E_{a2} = E_{b2} = 2.3495$ eV for the same coupled oscillator model. The cavity modes are much more positively detuned, and thus indices are greater than those in Fig. 2 due to Kramers-Kronig relations. The cavity parameters are taken as $E_{ph\_a} = 2.3610$ eV, $n_{eff\_a} = 3.45$, $E_{ph\_b} = 2.3660$ eV and $n_{eff\_b} = 3.42$. As the pump power increases to 0.4 P$_{th}$, a strong blue shift is observed in Fig. 3. We attribute this primarily to the excitonic blue shift from excitonic interactions. It is also worth noting that the blue shift after threshold is almost an order of magnitude stronger than ref. 28 of the main text, indicating a much stronger polariton polariton interaction. In the coupled oscillator model, we assume the cavity modes do not shift and the effective refractive indices do not change. The exciton energies here are $E_{a1} = E_{b1} = 2.3375$ eV and $E_{a2} = E_{b2} = 2.3515$ eV. The cavity detunings and effective indices are kept the same for the lower power and do not change for all the higher powers. For the pump power at P$_{th}$ and 1.4 P$_{th}$, the Rydberg energies are $E_{a1} = E_{b1} = 2.3420$ eV and $E_{a2} = E_{b2} = 2.3565$ eV.

**Supplementary Note 2: Comparison of Mott density and exciton density at threshold**

For an electron-hole system described by classical Boltzmann statistics with Debye–Huckel screening, at finite temperature, the Mott density $n_M$, where the electron-hole plasma starts to exist and the excitons stop to exist as individual quasi-particles, is approximated by (6)

$$n_{M-3D} = (1.19)^2 \frac{k_B T}{2 a_B^3 R_y^*}$$

Where $k_B$ is the Boltzmann constant, $T$ is sample temperature, $L$ is the perovskite layer thickness, $a_B$ is the Bohr radius, and $R_y^*$ is Rydberg constant, i.e. the exciton binding energy. Such a formula is derived by considering the fact that when Debye–Huckel screening length falls below a critical value, the exction's Yukawa potential no longer has a bound state in three-dimensional systems. The estimated 3D Mott Density at 55K is $\approx 1 \times 10^{17}$ cm$^{-3}$, when $a_B \approx 3.5$ nm and $R_y^* \approx 25$ meV (7);

The effective generated exciton density inside perovskite can be estimated from pump laser raw pulse energy measurement, if one knows the reflective of the top Ag mirror and absorption of the perovskite layer, using (8)

$$n_{ex-3D} = (1-R)(1-A_{ag})AI_p / (f_p \hbar \omega \pi r^2 t)$$

where $R$ is the reflectivity of the cavity's top mirror, $t$ is sample thickness, $A$ is the absorbance of the perovskite crystal, $I_p$ is the average pump power, $f_p$ is the pulsed laser repetition rate, $\hbar\omega$ is the pump photon energy and $r$ is the pump beam radius.

Note here such an equation overestimates the exciton density because it does not consider the exciton quantum yield (i. e. what fraction of the absorbed photons turn into excitons). $I_p$ needs to be normalized with the corresponding quantum yield. The exciton quantum yield can be greatly decreased by the cavity fabrication process, mainly due to possible quenching centers at the interfaces of the layered structures. The external PL quantum yield of the sample is a good indicator of the exciton quantum yield. The highest PL quantum yield achieved in our CVD grown perovskites is ~50%. The quantum yield of our perovskite samples drops to ~5% with Al seeding layers (Fig. 2 and Fig. 3), and can be improved to ~40% with a 10 nm $Sb_2O_3$ seeding layer. We have measured 8 samples, 3 of which are from the Al seeding layer experiments (one typical example is described in the main text) and 5 from $Sb_2O_3$ seeding layer (one typical example is described in Fig. S6). By taking the statistics of all the measured samples, the normalized $I_p$ is averaged to around 800 uW (12 MW/cm² effective pump peak density or 2.4 µJ/cm² pulse energy, after taking the Ag mirror reflectivity and absorption into consideration) for our pulsed femto-second laser with a repetition rate of 80 MHz. Given the experimentally determined top Ag mirror Reflectivity R ≈ 96% and Absorption $A_{ag}$ ≈ 24% and perovskite layer absorption A ≈ 50%, the estimated upper limit of the average 3D exciton density at threshold is ▯ $3.7 \times 10^{15} cm^{-3}$, which is more than 25 times smaller than the Mott density at our experimental temperature. This estimation further justifies our claim of exciton polariton condensate in a perovskite microcavity.

## Supplementary Note 3: Mode competition $LP_a$ and $LP_b$ polariton modes

We elaborate here on the mode competition discussed in the main text. Strong exciton-exction interactions are well known as the driving force to realize polariton condensation. Larger exciton fractions of polariton (or in another word, smaller photon fractions of polariton) will certainly facilities the polariton-polariton interaction. Because of the more positive detuning of the polariton LP$_b$ mode, it has a smaller photon fraction, a stronger exciton-exciton interaction, and thus a kinetically favorable gain coefficient as the pump power initially increases. As the pump power increase continues, some complex internal nonlinear effects, such as inhomogeneous gain saturation, start to kick in and the LP$_b$ mode gain become smaller than the gain of the thermal dynamically lower energy LP$_a$ mode. Thus, the crosstalk between the two orthogonal lower polariton modes of LP$_a$ and LP$_b$ can be simply viewed as a competition between the kinetically favorable metastable state LP$_b$ and the thermal dynamically favorable state LP$_a$.

Condensate behavior is normally modelled by a dissipative Gross-Pitaevskii equation (GPE) (9):

$$i\frac{\partial \psi(r,t)}{\partial t} = [-\frac{\hbar \nabla^2}{2m_{LP}} + \frac{i}{2}(R(g, g_R, n_R, \Delta)n_R - \gamma_{LP}) + g|\psi(r,t)|^2 + g_R n_R]\psi(r,t)$$

$$\frac{\partial n_R(t)}{\partial t} = P(t) - (\gamma_R + R(g, g_R, n_R, \Delta)|\psi(r,t)|^2)n_R + D\nabla^2 n_R$$

In this simple phenomenological model, the polariton dispersion is approximated by a parabolic shape with effective mass $m_{LP}$ and loss rate $\gamma_{LP}$. The strengths of polariton-polariton interactions and polariton-reservoir interactions are described by the coupling constants g and $g_R$.
The amplification rate of the condensate due to exciton-exciton interactions is phenomenologically described as $R(g, g_R, n_R, \Delta)$, where the polariton R is a function of interaction strengths, reservoir density $n_R$ and detuning $\Delta$. The lower polariton condensate evolution is also coupled to the density of the reservoir $n_R$.

In practical experiments, for the single crystal of the perovskite samples, the inhomogeneity is generally very small. For small pump spot sizes of ~ 4 μm, the distribution can also be approximated to be uniform. Thus one can further simplify the model by ignoring the spatial dependence of both lower polariton and exciton reservoirs so that the equations can be written as:

$$i\frac{d\psi(t)}{dt} = [\frac{i}{2}(R(g, g_R, n_R, \Delta)n_R - \gamma_{LP}) + g|\psi(t)|^2 + g_R n_R]\psi(t)$$

$$\frac{dn_R(t)}{dt} = P(t) - (\gamma_R + R(g, g_R, n_R, \Delta)|\psi(t)|^2)n_R$$

Where the polariton density wave function is written as: $\psi(t) = e^{-i\mu t}\psi_0$ and the chemical potential blue shift of the system due to the exciton-exciton interactions is given by: $\mu = g|\psi(t)|^2 + g_R n_R$.

When no condensation is present, $\psi = 0$, and the density of exciton reservoir $n_R = \dfrac{P}{\gamma_R}$. If a steady state condensate is present, then $\dfrac{d\psi_i(t)}{dt} = 0, \dfrac{dn_R(t)}{dt} = 0, |\psi(t)| \geq 0$. And the threshold pump density is given by $P_{th} = \dfrac{\gamma_R \gamma_{LP}}{R(g, g_R, n_R, \Delta)}$. For a given reservoir and associated $\gamma_R$, the condensation threshold thus only depends on experimental controlled parameter $\gamma_{LP}$ and the system detuning related function $R(g, g_R, n_R, \Delta)$.

For the system described in the manuscript, one can write a similar phenomenological GPE model:

$$i\frac{d\psi_a(t)}{dt} = [\frac{i}{2}(R_a(g_a, g_{aR}, n_R, \Delta)n_R - \gamma_{LPa}) + g_a |\psi_a(t)|^2 + g_{aR} n_R]\psi_a(t)$$

$$i\frac{d\psi_b(t)}{dt} = [\frac{i}{2}(R_b(g_b, g_{bR}, n_R, \Delta)n_R - \gamma_{LPb}) + g_b |\psi_b(t)|^2 + g_{bR} n_R]\psi_b(t)$$

$$\frac{dn_R(t)}{dt} = P(t) - (\gamma_R + R_a(g_a, g_{aR}, n_R, \Delta)|\psi_a(t)|^2 + R_b(g_b, g_{bR}, n_R, \Delta)|\psi_b(t)|^2) n_R$$

When $\psi_b = 0$ $\psi_a \neq 0$ the system will condensate in the LP$_a$ polariton mode at the pump power $P_{ath} = \dfrac{\gamma_R \gamma_{LPa}}{R_a(g, g_R, n_R, \Delta)}$; and vice versa when $\psi_a = 0$ $\psi_b \neq 0$ the system will condensate to polariton LP$_b$ mode at the pump power $P_{bth} = \dfrac{\gamma_R \gamma_{LPb}}{R_b(g, g_R, n_R, \Delta)}$. Given the highly nonlinear and complex nature of $R(g, g_R, n_R, \Delta)$, it is challenging to quantitatively describe this function. Qualitatively, however if for $n_R < n_{th}$, $R_b(g_b, g_{bR}, n_R, \Delta)n_R - \gamma_{LPb} > R_a(g_a, g_{aR}, n_R, \Delta)n_R - \gamma_{LPa}$, as an outcome the population of lower polariton modes of LP$_b$ will increase faster than LP$_a$ and $P_{ath} > P_{bth}$ leads to the fact that LP$_b$ will condensate first; if at $n_R \sim n_{th}$ and $n_R > n_{th}$, $\dfrac{\gamma_{LPb}}{R_b(g, g_R, n_R, \Delta)} > \dfrac{\gamma_{LPa}}{R_a(g, g_R, n_R, \Delta)}$ and $R_b(g_b, g_{bR}, n_R, \Delta)n_R \approx \gamma_{LPb}$, then LP$_a$ will condensate while LP$_b$ will roughly remain a stable population.

## Supplementary Note 4 : Calculation of polariton interactions

The polariton interaction is ultimately due to Coulomb interactions between electrons and holes in the matter. The bare Coulomb interaction is

$$\frac{1}{2\Omega}\sum_{\vec{k},\vec{p},\vec{q}}V_{\vec{q}}[a^{\dagger}_{c,\vec{k}+\vec{q}}a^{\dagger}_{c,\vec{p}-\vec{q}}a_{c,\vec{p}}a_{c,\vec{k}}+a^{\dagger}_{v,\vec{k}+\vec{q}}a^{\dagger}_{v,\vec{p}-\vec{q}}a_{v,\vec{p}}a_{v,\vec{k}}-2a^{\dagger}_{c,\vec{k}+\vec{q}}a_{v,\vec{p}}a^{\dagger}_{v,\vec{p}-\vec{q}}a_{c,\vec{k}}]\;,\text{ where }a^{\dagger}_{c(v),\vec{k}}$$

and $a_{c(v),\vec{k}}$ are conduction (valence) band electron creation and annihilation operators, $\Omega$ is the three-dimensional system volume, and $V_{\vec{q}}=4\pi e^2/\varepsilon q^2$ is the three-dimensional Coulomb interaction. By mapping the conduction/valence band electrons to excitons(10), we obtain the two-body exciton-exciton interaction matrix element $M_{\vec{k},\vec{p},\vec{q}}=2\sum_{\vec{k},\vec{p}}V_{\vec{k}-\vec{p}}(|\varphi_k|^2\varphi^*_k\varphi_p-|\varphi_k|^2|\varphi_p|^2)$, where $\varphi_k$ is the exciton wave function. Note that we ignore the long-range dipole-dipole terms in the direct Coulomb interaction channel and consider only short-range exchange contributions at zero center-of-mass momentum. These approximations yield $U_{ex}\approx 27.2R^*_y a^3_B$ for 1s excitons and $U_{ex}\approx -68.5R^*_y a^3_B$ for 2s excitons, where the effective Bohr radius $a_B\approx 3.5$ nm, and effective Rydberg energy $R^*_y\approx 25$ meV. The attractive interaction for 2s excitons is due to the dominant negative electron-electron exchange contributions with a larger exciton size(11). To account for these short-range exciton-exciton interactions, we add a term $\frac{1}{2}U_{ex}n^2_{ex}$ to the noninteracting exciton-photon energy functional, and this term leads to the energy blue shift of polaritons, $U_{pol}n_{pol}$, which is a linear function of polariton density in the low density limit. For the three-branch 1s exciton-2s exciton-photon model, we have to numerically compute the polariton interaction strength based on the exciton fraction. In order to make a comparison with the experimental observed blue shift, we crudely estimate that the exciton density is 50% of the total excitation density (12) and that all these excitons contribute to the hybridized polariton. By assuming that 3D excitons are uniformly stacked by 2D excitons, we can convert the pumping power to an effective 2D polariton density for *a* and *b* polarization at a given exciton/polariton ratio.

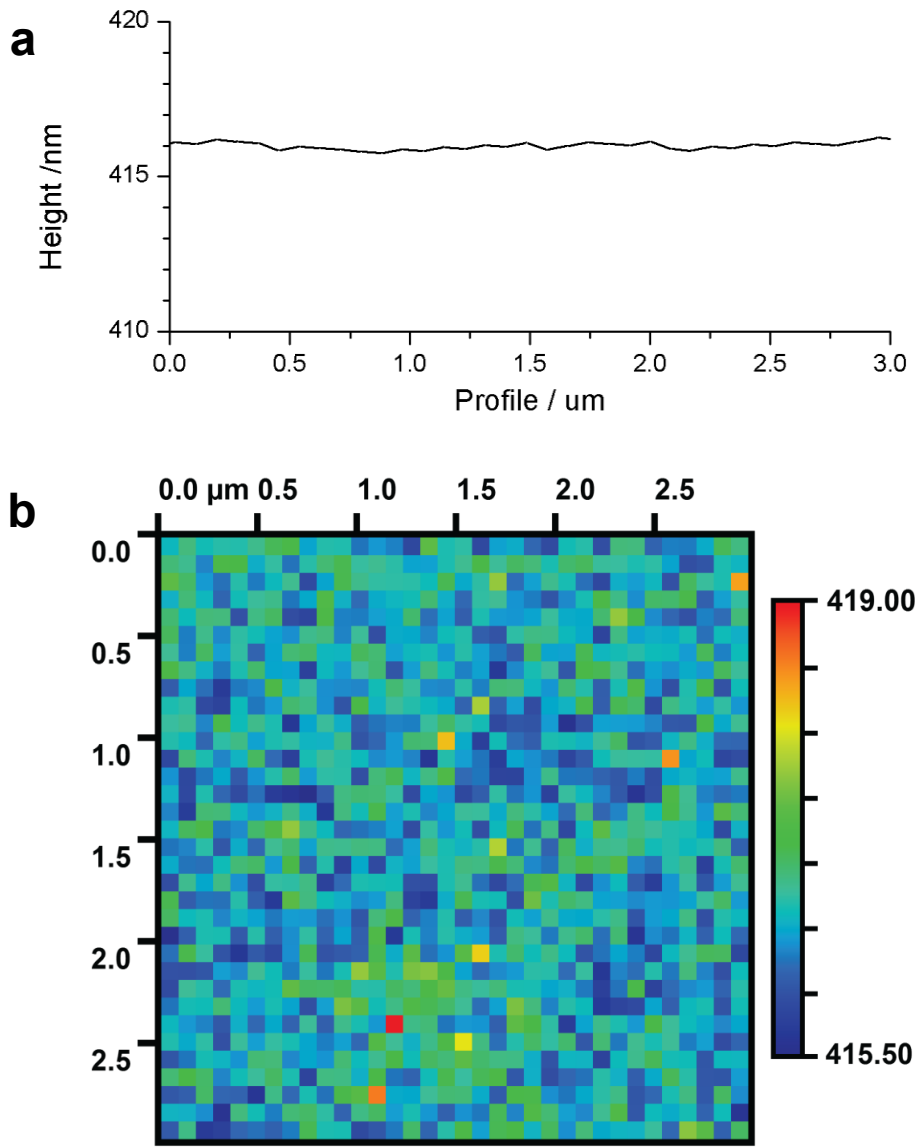

**Figure S1** Thickness and flatness charaterization of perovskite. (a) Non-contact AFM measured cross-section line profile suggests total thickness of the device ~416 nm (Fig. 1b) and truely atomic flat interface within the pump laser spot size. (b) A zoom-in area of 3um x 3um AFM image of the same sample in Fig. 1 and Fig. 3.

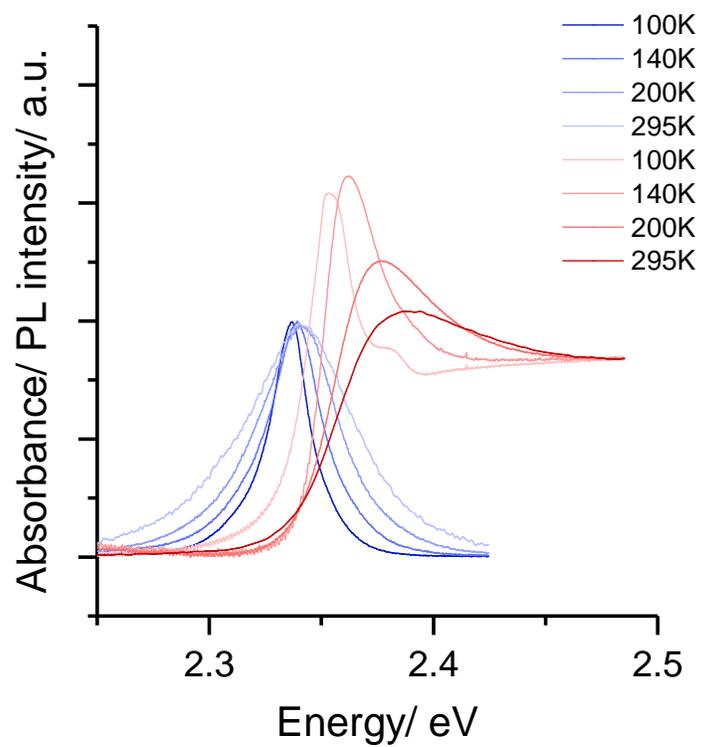

**Figure S2.** Temperature dependent absorption (red curves) /PL (blue curves) spectra of CVD CsPbBr$_3$. As the temperature decreases, absorption peak shows red shift.

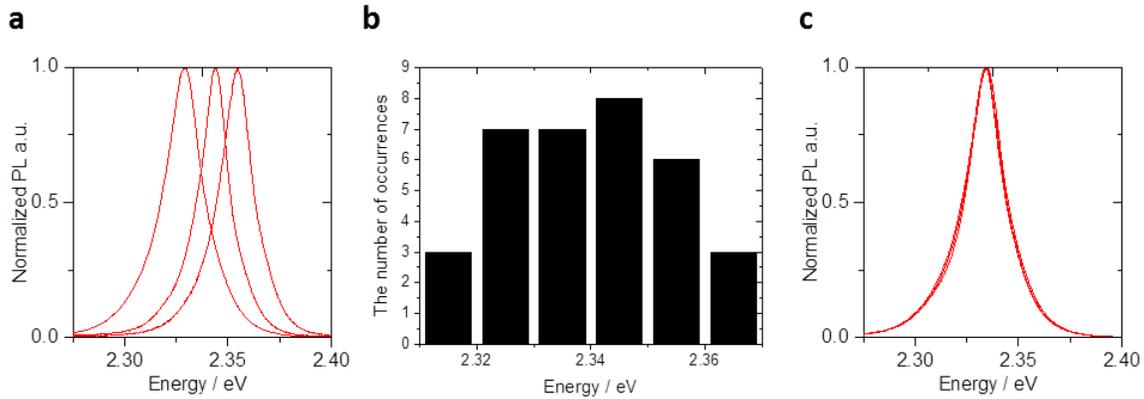

**Figure S3**. PL sample to sample variation at 55K. **a.** Typical PL peaks at different microplates grown on the same mica substrate. **b.** PL peak location statistics. **c.** Multiple PL measurements of the same microplate at 3 different locations (~5μm separation) show almost identical PL spectra and only a small amount spatial disorder within the single crystal microplates.

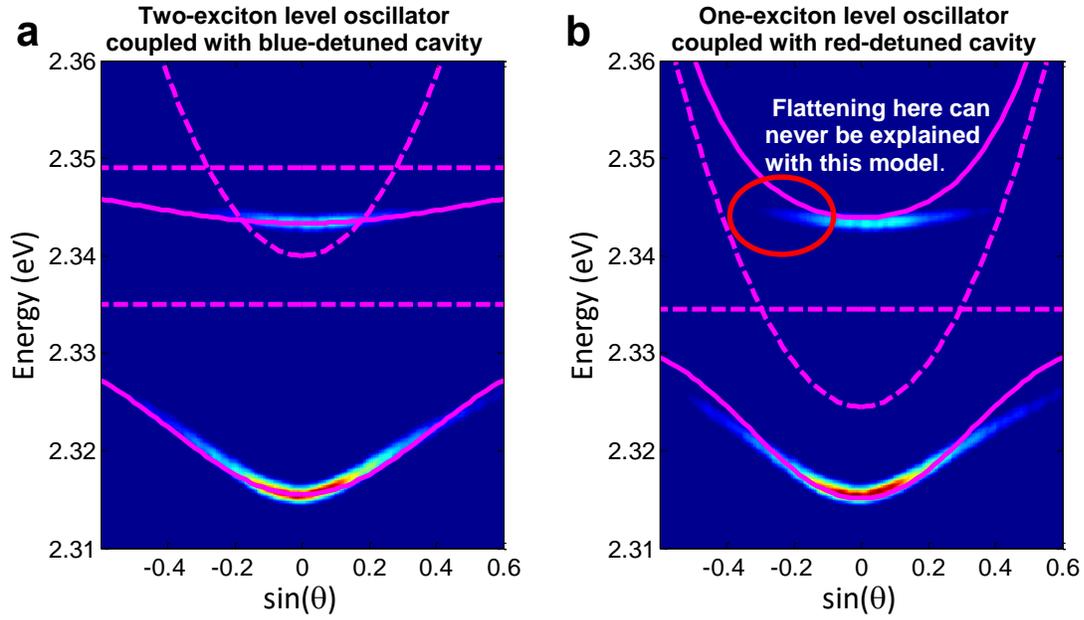

**Figure S4.** Comparison of coupled oscillator models for polariton dispersion with two-exciton levels and one-exciton level. **a**, The two-exciton level fit to the k-space PL map along crystal axis a (as in Fig. 2a in the main text). **b**, The one-exciton level fit to the k-space PL map along crystal axis a. In these two maps, the exciton energy level and photonic cavity mode before strong coupling (dashed line) and the fitted polariton dispersion (solid line) are overlaid with the PL map. Clearly, as emphasized by the red circle in panel **b**. the high angle flattening can never be explained with a one exciton polariton model. In contrast an accurate fit to the dispersion is obtained within a two-exciton level model. The two levels are in our case the two lowest members of a Rydberg series.

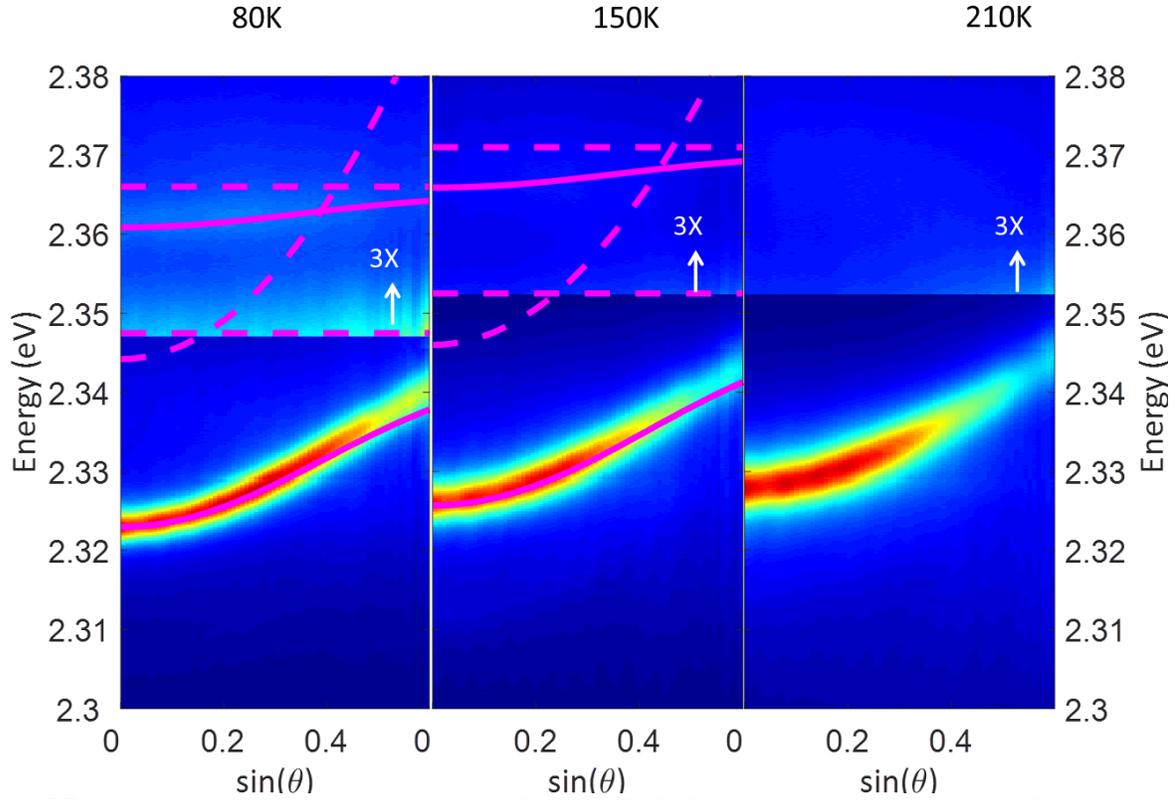

**Figure S5**, Temperature-dependent k-space PL of a CsPbBr$_3$ perovskite cavity. The PL collected here is based on a new sample with a polarization along the **b** axis. As the temperature increases from 80K to 150K, and to 210K, the lower branch blue-shifts while the middle branch becomes smeared out. When temperature is larger than 150K, only the lower branch is visible. The PL dispersion fit is based on a three coupled oscillators. The purple dashed lines represent the 1s and 2s exciton energies ($E_1$ = 2.347 eV and $E_2$ = 2.366 eV at 80K, $E_1$ = 2.352 eV and $E_2$ = 2.371 eV at 150K), respectively. The dashed curve represents the cavity dispersion, where $E_{cav}$ = 2.344 eV at 85K and 2.346 eV at 150K. The solid purple curves represent the dispersion of polariton branches, consistent with the observed PL emission. The Rabi splittings fits are $\Omega_1$ =(44 ± 1) meV and $\Omega_2$ =(25 ± 1) meV. The excitonic energies and Rabi splittings have minor variations, but are experimentally consistent with Fig. 2 in main text. Note that the PL emission above $E_1$ is magnified by three times for better visualization.

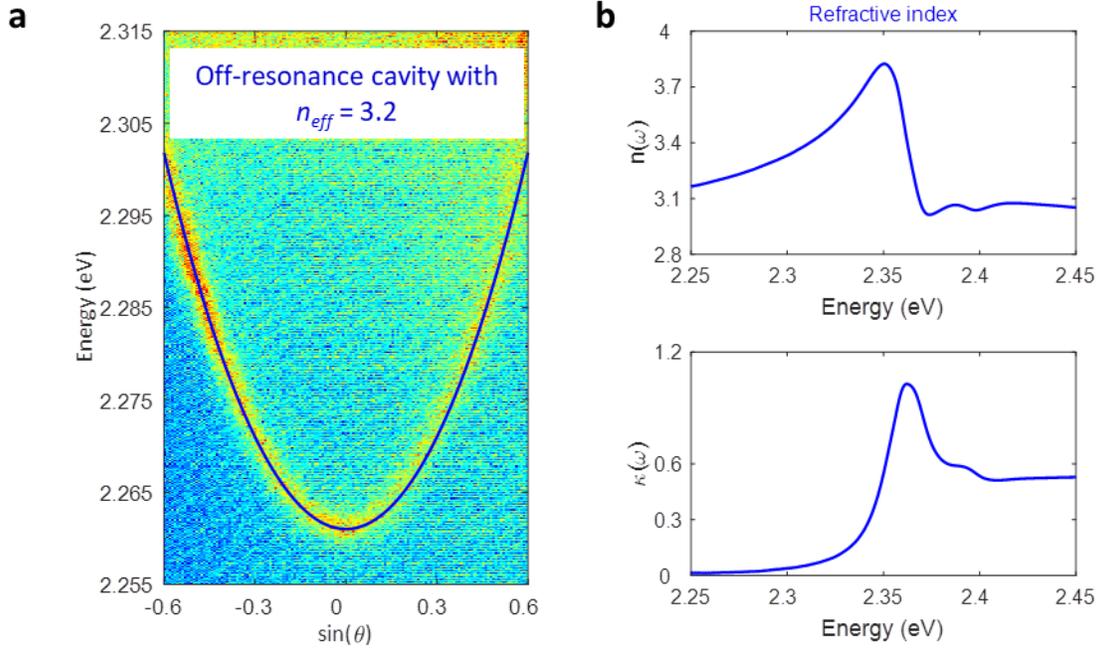

**Figure S6.** Off-resonance cavity dispersion and refractive index. **a,** Based on the off-resonance cavity dispersion, an effective refractive index of $n_{eff} = 3.2$ is obtained, and thus a background refractive index of 2.6 is derived. **b,** Based on the absorption spectrum and Kramers-Kronig relations, the real and imaginary parts of the refractive index of $CsPbBr_3$ are obtained. Based on the data, Fig 2 implies an effective index of 3.8, and 3.75 for the $n_{eff\_a}$ and $n_{eff\_b}$; Fig 3 implies effective indices of 3.45, and 3.42 for the $n_{eff\_a}$ and $n_{eff\_b}$.

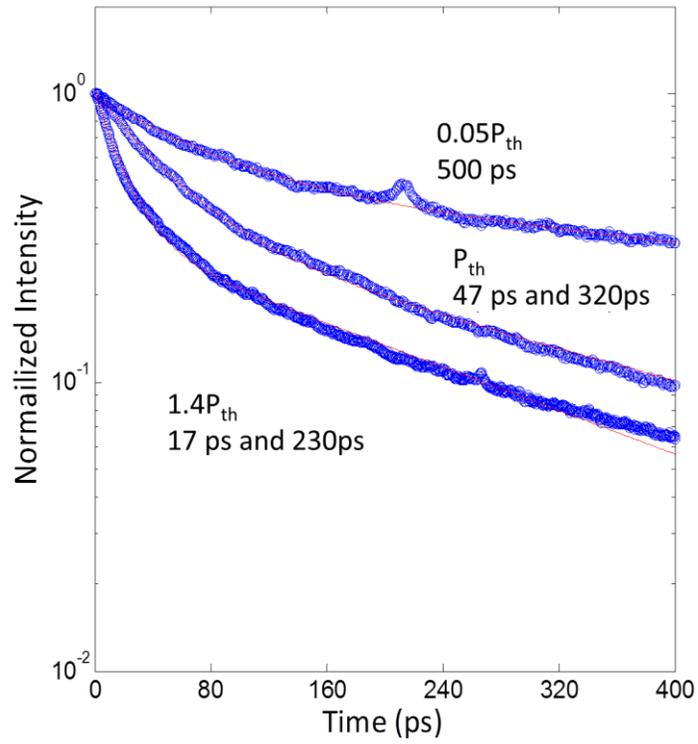

**Figure S7.** Polariton spectrally-integrated temporal response dependence of pump power measured by streak camera. Below threshold, a single exponential decay is observed, while above threshold at $P=1.4P_{th}$, bi-exponential decay appears with the fast decay component as fast as 17 ps. All the measurements share the same settings and the small peak at $P=0.05\ P_{th}$ ~200 ps is random detector noise.

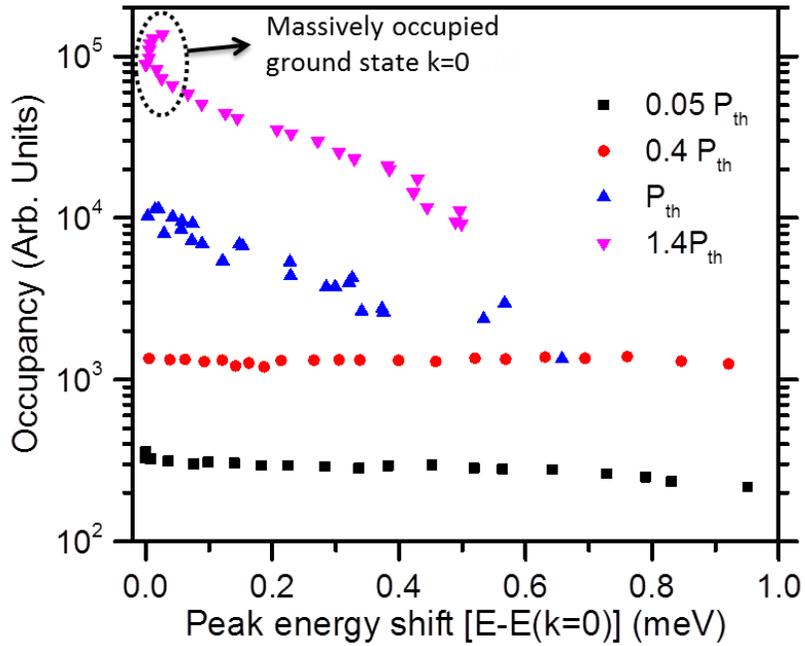

**Figure S8.** Polariton occupancy in ground- (k=0) and excited-state levels. Polariton occupancy distribution of $LP_a$ branches, obtained from data in Fig. 3, at 0.05 $P_{th}$, 0.4$P_{th}$, $P_{th}$ and 1.4$P_{th}$ pump power. Below the excitation threshold, the $LP_a$ polariton occupation follows Boltzmann-like exponential decay; while above threshold, the ground state of $LP_a$ k=0 becomes massively occupied, which is a typical signature of BEC state formation (13, 14).

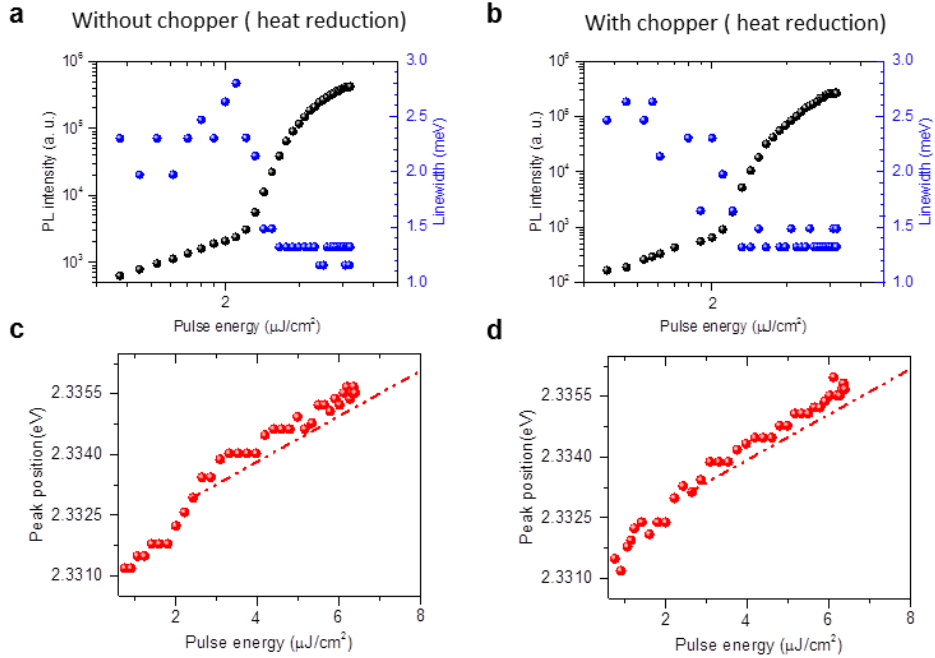

**Figure S9**. Exciton-polariton condensate behavior with heat reduction and without heat reduction at 55 K. (a), (b) Log-log plot of integrated PL intensity of $LP_a$ mode at $\theta = 0°$ and full-width at half-maximum (FWHM) of $LP_a$ mode at $\theta = 0°$ versus pump power. Nonlinearity and linewidth narrowing of the polariton mode is clearly observed as the excitation intensity exceeds the almost identical condensation threshold. (c) (d) PL polariton peak position of $LP_a$ mode (red dot) at $\theta = 0°$ versus pump power. A similar amount of strong blue shift is observed due to the strong exciton interactions. The theory predicted blue shift contributed from 1s-exciton-resulted polariton interaction is plot in red dot-dot-dash line for guidance.

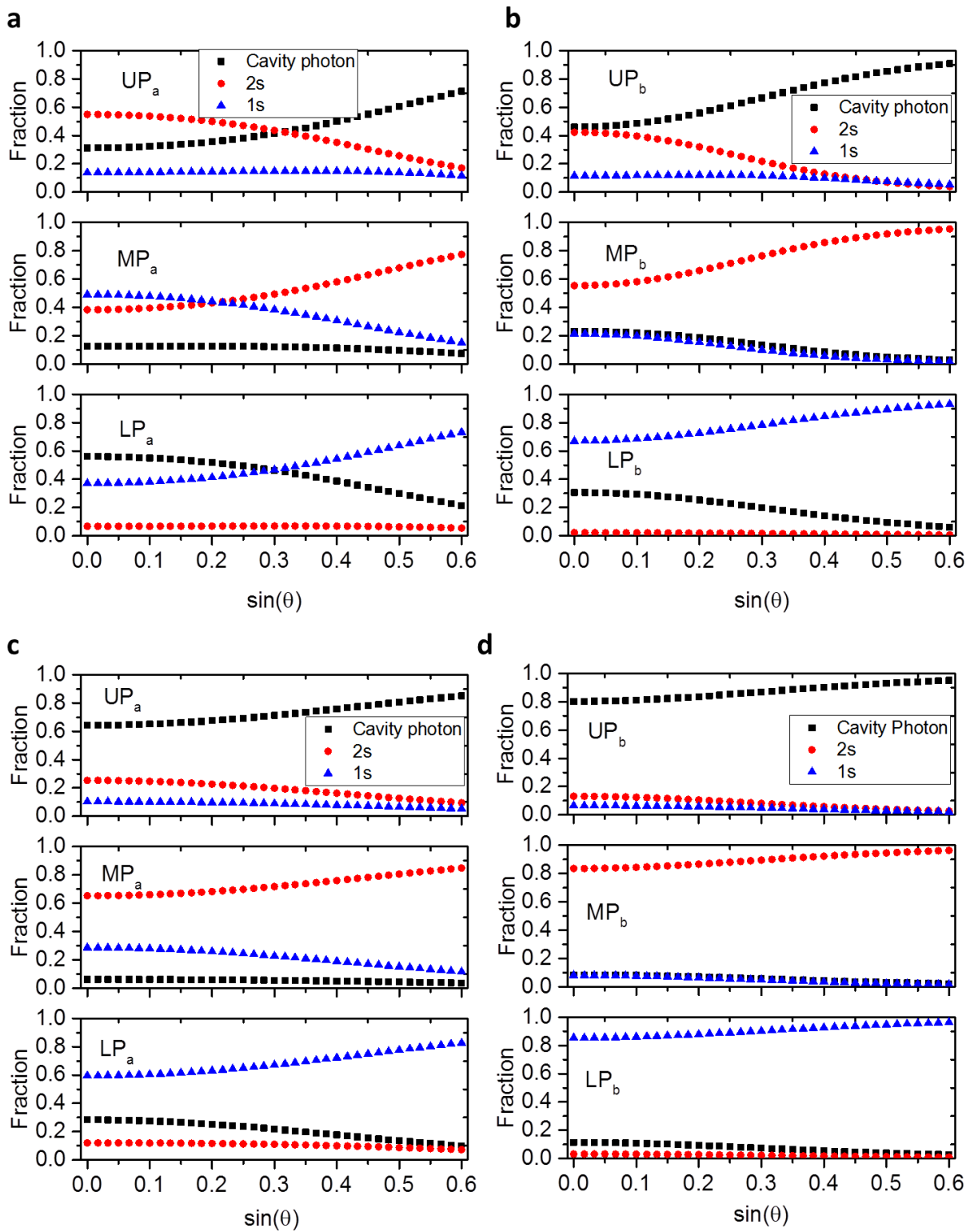

**Figure S10**, Hopfield coefficients for the polariton dispersions in the main text. **a** and **b**, Hopfield coefficients of 1s, 2s excitons and cavity photons for the polariton dispersions with polarizations

along *a* and *b* axis in the main text Fig. 2**a** and **2b**. **c** and **d** Hopfield coefficients of 1s, 2s excitons and cavity photons for the polariton dispersions before condensation with polarizations along *a* and *b* axis in the main text Fig. 3**a** (the first panel).

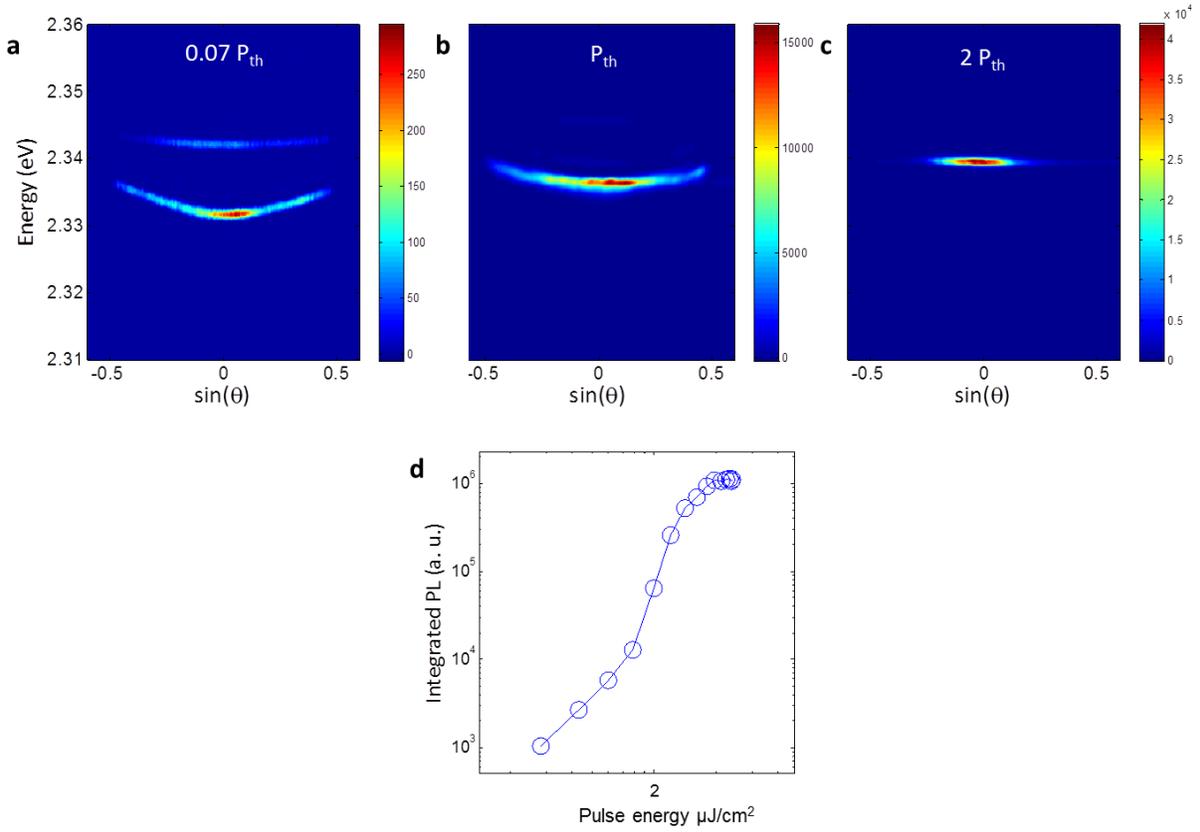

**Figure S11**. Exciton-polariton condensate behavior at 55K from a typical $Sb_2O_3$ seeding layer sample. Polarization selective k-space power dependent angle resolved PL map of $LP_b$ mode taken at **(a)** $0.07P_{th}$, **(b)** $P_{th}$, and **(c)** $2P_{th}$ (from left to right). The excitation is 460 nm light polarized along the diagonal of ***a*** and ***b*** axes. The sample is slightly thicker and more negative detuned than in Fig. 3 of the main text. The condensate in $LP_b$ mode is due to the more negative detuning. **(d)** Log-log plot of integrated PL intensity of $LP_b$ mode at $\theta = 0°$ versus pump power. The lower threshold is a result of better preserved PL quantum yield during fabrication process.

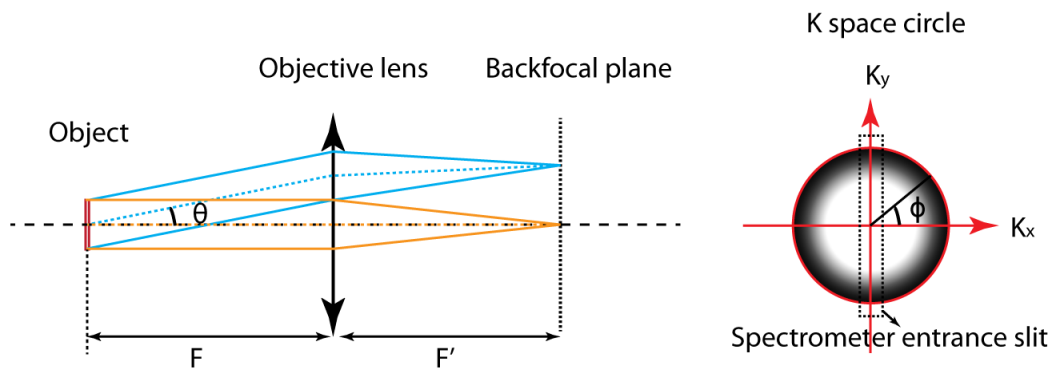

**Figure S12.** The schematic of back focal plane (BFP) of an objective lens. The F and F' represent the front and back focal lengths of the lens, respectively. As illustrated by the blue and orange lines, the rays emitted from the object to the same angular orientation lie onto the same lateral coordinate at the BFP. The spectrometer entrance slit selects one line of the k-space circle for the spectroscopic imaging. The k-space spectroscopy is based on this back focal plane imaging. Each point in the back focal plane (Fourier plane) is the superposition of photon fields emitted by the source along a given direction $(\theta, \phi)$ where $\theta$ is the emission angle with respect to the normal to the microcavity plane and $\phi$ is the azimuth angle. This Fourier plane is imaged through the narrow entrance slit of the spectrometer and finally onto the 2D EMCCD array. The slit transmits one line along the diameter of the Fourier plane; each point of this line along one dimension of this EMCCD array represents the emission angle, and then the grating in the spectrometer disperses the wavelength of this line along the other dimension of the EMCCD array. The image shows k-space map with the angular dependence for both reflectivity and PL.

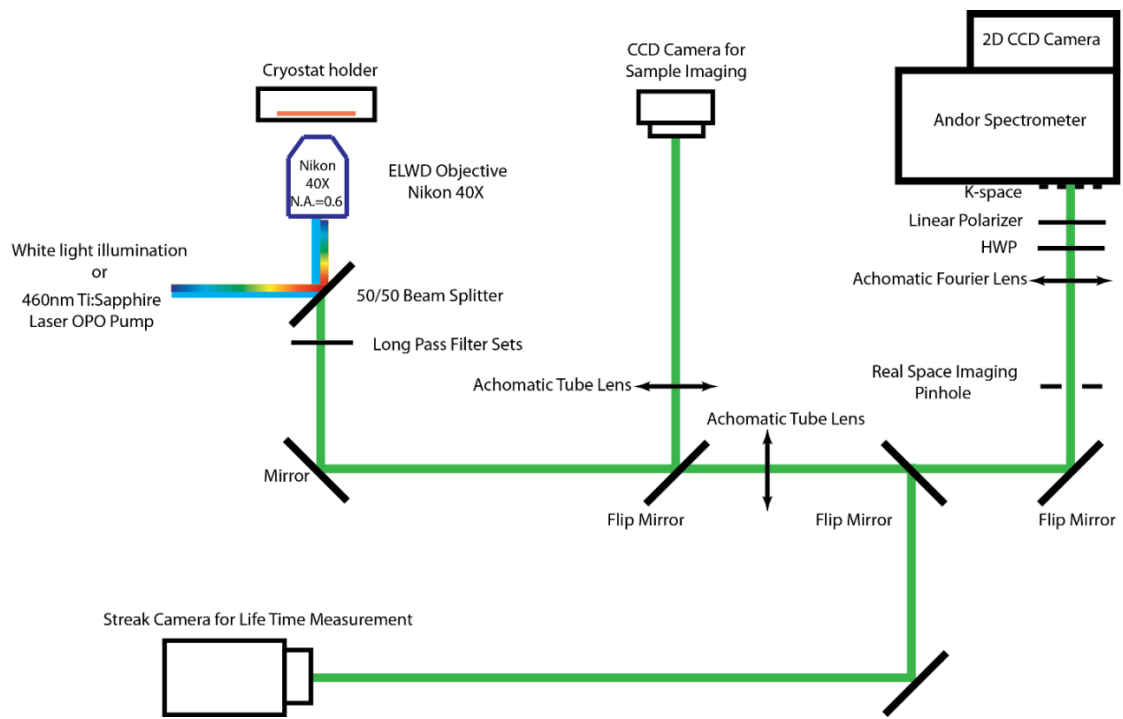

**Figure S13.** Schematic of the home-built setup for laser characterization, lifetime measurement, and polarization-dependent k-space photoluminescence (PL) spectroscopy.